\documentstyle[11pt,aaspp4,epsf,ifthen]{article}
\newboolean{figs}\setboolean{figs}{true}

\def\hst{{\it HST}}
\def\etal{{\it et al.}}
\def\Msun{M_{\odot}}

\def\subr #1{_{{\rm #1}}}
\def\supr #1{^{{\rm #1}}}
\def\disp #1{\langle v^2\subr{#1} \rangle}
\def\vv #1{v^2_{#1}}
\def\vvr #1{v^2\subr{#1}}

\def\psubm{p_{\rm m}}

\def\minspt{\farcm}
\def\secspt{\farcs}
\def\spt{\fs}

\begin{document}

\title{Mass Segregation and Equipartition of Energy in Two \nl
Globular Clusters with Central Density Cusps\footnote{Based on
observations with the NASA/ESA {\it Hubble Space Telescope}, obtained
at the Space Telescope Science Institute, which is operated by AURA,
Inc., under NASA contract NAS 5-26555.}}
\author{Craig Sosin}
\affil{Astronomy Department, University of California, Berkeley, CA
94720-3411}
\affil{E-mail: csosin@astro.berkeley.edu}

\begin{abstract}
We begin by presenting the analysis of a set of deep $B$- and $V$-band
images of the central density cusp of the globular cluster M30 (NGC
7099), taken with the Faint Object Camera aboard the {\it Hubble Space
Telescope.}  These images are the first to resolve lower-mass
main-sequence stars in the cluster's central $10''.$ From the
positions of individual stars, we measure an improved position for the
cluster center; this new position is $2\secspt6$ from the previously
known position.  We find no evidence of a ``flat'',
constant-surface-density core; however, the data do not rule out the
presence of a core of radius up to $1\secspt9$ (95\% confidence
level).  We measure a logarithmic cusp slope ($d \log \sigma / d \log
r$) of $-0.76 \pm 0.07$ (1-sigma) for stars with masses between $0.69$
and $0.76 \Msun$, and $-0.82 \pm 0.11$ for stars with masses between
$0.57$ and $0.69 \Msun$.  We also compare the overall mass function
(MF) of the cluster cusp with the MF of a field at $r = 4\minspt6$
(near the cluster half-mass radius).  The observed degree of mass
segregation is well matched by the predictions of an isotropic,
multimass King model.

We then use the Jeans equation to compare the structure of M30 with
that of M15, another cusped cluster, using data from this and a
previous paper.  We find that M30 is very close to achieving
equipartition of energy between stellar species, at least over the
observed range in mass and radius, while M15 is not.  This difference
may be a result of the longer relaxation time in the observed field in
M15.  The data also suggest that the degree of mass segregation within
the two cluster cusps is smaller than one would expect from the
measurements at larger radius.  If so, this phenomenon might be the
result of gravothermal oscillations, of centrally-concentrated
populations of binaries, or of a $\sim 10^3 \Msun$ black hole in one
or more clusters.
\end{abstract}

\keywords{globular clusters---stellar systems (kinematics, dynamics)}

\clearpage

\section{Introduction}

Theoretical arguments, computational simulations, and high-resolution
observations have all shown that the cores of high-concentration
globular clusters collapse, on timescales comparable to the Hubble
time.  The result of this collapse is a power-law density cusp,
believed to be supported against further collapse by the transfer of
energy out of binary stars through stellar encounters (see the reviews
in Djorgovski\markcite{ivanfest} \& Meylan 1993a and Hut\markcite{iau}
\& Makino 1996).  With the observation of density cusps in $\sim 20\%$
of Galactic globulars (Djorgovski\markcite{djk86} \& King 1986,
Lugger\markcite{lug95} \etal\ 1995), core collapse has become the
standard model of cluster dynamical evolution; alternative
explanations for the existence of the cusps, such as the presence of a
central black hole (Bahcall \& Wolf \markcite{bw76}1976,
\markcite{bw77}1977) are now less favored.  Recently, the repair of
the {\it Hubble Space Telescope (HST)} has allowed faint stars in
cluster cusps to be observed for the first time, and has made detailed
tests of these models possible.

In a previous paper (Sosin \& King \markcite{paper1}1997, hereafter
Paper I), we presented our analysis of several \hst/Faint Object
Camera (FOC) images of the central cusp of the cluster M15.  Here, we
begin by examining a similar set of images of the cluster M30 (NGC
7099).  As before, our primary goals are to investigate the
distribution of stars of different mass, and to use those
distributions to test theories of the structure and evolution of
globular clusters.  After presenting the M30 data, we compare the two
clusters, and suggest some possible reasons for the differences
between them that we observe.

Like M15, M30 is a cusped---presumably post-core-collapse---cluster,
with low metal abundance ([Fe/H] $= -2.1$) and advanced age.
\hst/WFPC2 imaging has recently shown that M15 and M30 have nearly
identical global mass functions (Piotto, Cool, \& King
\markcite{pck97}1997), suggesting that the pair were formed in very
similar circumstances.  However, M30 has only about one-third the mass
of M15, and a shorter half-mass relaxation time (Djorgovski
\markcite{table1}1993).

\section{The M30 Data}

\subsection{Observations}

M30 was observed by \hst\ in December 1994, with the COSTAR optical
correction system in place.  Two $\sim 7'' \times 7''$ FOC fields were
observed near the cluster center on 5 December, and an additional FOC
field at a radius of $21''$ was observed two days later.  This latter
field unfortunately includes a bright giant that saturated a large
portion of the FOC detector; that fact plus the lower stellar density
at $r=21''$ imply that there are too few stars in the field to measure
a meaningful luminosity function.  For the rest of this paper we
consider only the two inner fields.

Each of these two fields was observed through the FOC equivalents of
$B$ and $V$ filters (F430W and F480LP, respectively) for a total of
$\sim 4000$ seconds.  Their combined $V$ image is shown in Figure 1.
The first field was intended to be positioned so that the cluster
center would be near its midpoint, however, errors in the previously
known position of the center led to the center appearing near the left
edge of the image.  The second field is about $7''$ from the first;
the two overlap by only a few pixels along one edge.  Together, the
two span a radial range from the cluster center out to $r \simeq
12''$, far enough out that the images show the cusp's entire radial
extent.

The FOC detector saturates at high count rates, rather than at high
numbers of counts, so saturation of stellar images is often a problem
in FOC images.  The large, crescent-shaped ``objects'' in Fig.\ 1 are
saturated giants; their odd shape is a result of hysteresis effects in
the detector.  In addition, many of the brightest main-sequence stars
in the images have nonlinear pixels in their centers, especially in
the $B$ image.  In Paper I, the extreme crowding of saturated stars
made the $B$ images of the center of M15 difficult to work with, but
here, both the $B$ and $V$ images of M30 are usable.

The M30 images were also taken shortly after the FOC was turned on,
after having been off for several days.  Because the detector was
still warming up, the plate scale changed slowly over the course of
the observations.  This effect is especially pronounced for the $V$
image of the second field (which was actually the first to be taken),
in which stars are somewhat distorted near the image corners.
However, since this distortion appears in the same parts of the images
in which the FOC's flatfielding and geometric correction are poor even
under normal conditions, it affects the photometric accuracy very
little.

\subsection{Data Reduction}

We used the same procedure as in Paper I to obtain positions and
magnitudes of stars in the two M30 fields.  We refer the reader to the
previous paper for an in-depth description of the method.
Essentially, we run DAOPHOT and ALLSTAR (Stetson \markcite{pbs87}1987,
\markcite{pbs92}1992) on the image twice; the results from the first
pass are used to subtract features of the FOC point-spread function
that are easily confused with faint stars (e.g., intersections of
diffraction spikes with Airy rings).  The second pass, which uses this
``ring-subtracted'' image as input, can then reliably find stars at
fainter magnitudes.  The lower stellar density in M30 made the third
pass from Paper I, which was used to split close stellar blends,
unnecessary.

We calibrated our photometry to standard FOC instrumental magnitudes
using photometric parameters provided by the STScI pipeline, and then
transformed to Johnson $B$ and $V$, as described in Paper I.  Since
the metal abundances of M15 and M30 are so similar, we used the same
color equations as in the previous paper (King, Anderson, \& Sosin
\markcite{kas94}1994).

As in Paper I, the points on the $(V,B-V)$ diagram came out somewhat
bluer and/or fainter than points on ground-based CMDs of the same
cluster.  We have no way of knowing whether the problem is in the FOC
calibration or in the color equations used to transform to $B$ and
$V$.  The discrepancy could be resolved in more than one way: for
example, the $B$ magnitudes could be made fainter by $0.15$ mag, or
the $V$ magnitudes could be made brighter by $0.12$ mag.
Alternatively, both magnitudes could change by some larger quantity.
Since the stars being observed are fainter than any observed
previously in the cusp, we have no other way to calibrate the
photometry, and must accept the nominal FOC calibration as being
approximately correct.  (The mass functions to be measured later are
affected little by a shift of 0.1 mag.)

The CMD of the M30 cusp is shown in Figure 2.  The dashed line in the
figure is the main-sequence ridgeline observed from the ground (at
larger radius) by Bolte\markcite{bol87} (1987), while the solid line
is the ([Fe/H] $= -2.03,$ $t = 16$ Gyr) isochrone of Bergbusch \&
VandenBerg\markcite{BV92} (1992, hereafter BV92).  (Later, we will use
the mass--luminosity relation associated with this isochrone.)  The
observed $B$ magnitudes have been adjusted 0.15 mag fainter in order
to match the ground-based ridgeline.  The redward deviation of the FOC
points from the isochrone at the bright end is due to the onset of
nonlinearity in the $B$ image.  The $V$ magnitudes of the stars shown
are nearly unaffected by nonlinearity.

We then ran artificial-star experiments to determine the completeness
of the star-counting as a function of magnitude and position.  About
$10^4$ artificial stars were added to each of the two fields (64 stars
were added per experiment), with randomly selected $V$ magnitudes, and
$B$ magnitudes chosen to put the artificial star on the main-sequence
ridgeline.  The images were then re-reduced in the same automated way
as the originals, and the output lists were examined to see which
artificial stars were recovered.  In addition, we also recorded the
input and output magnitudes of any real neighbors of the added
artificial stars.

The counts of stars of moderate magnitude ($V \simeq 21$) are nearly
complete over the entire radial range.  The few stars that are not
detected at these magnitudes are missed because of the obvious bright
stars seen in Fig.\ 1.  The photometry becomes more and more
incomplete at fainter magnitudes, before giving out at $V \simeq 23$,
as faint stars become lost in the fluctuations of the background light
(from sky and bright-star halos).  A comparison of magnitudes of
nonrecovered artificial stars with their real-star neighbors revealed
that blends of stars of nearly equal magnitude are much less common
here than in the M15 images.  Thus, we do not need to worry about the
uncertainties in completeness corrections that would result from those
blends (see Paper I).

It is likely that some of the stars in Fig.\ 2 are binaries.  Binaries
consisting of a main-sequence (MS) star and a dark stellar remnant
(e.g., a neutron star or faint white dwarf) are indistinguishable from
single MS stars in the CMD; however, the combined mass of the two
binary components is considerably larger than the mass of a single
star with the same magnitude.  If the fraction of stars that are
binaries is large, then the analysis of the dynamics becomes more
difficult, since we then cannot rely on the mass-luminosity relation
given by the BV92 isochrone.  Binaries composed of two MS stars, on
the other hand, would appear above the MS ridgeline by up to $0.75$
mag.  Rubenstein \& Bailyn\markcite{rb97} (1997) have recently
observed a high ($> 15\%$) MS binary fraction in the cusp of NGC 6752,
a cluster whose dynamical parameters are similar to those of M30.
However, in M30, there does not appear to be a large number of nearly
equal-mass MS--MS binaries, at least in the $21 < V < 22$ range where
they would be easily distinguished from single stars (see Fig.\ 2).
But, the photometric precision of the FOC is not really sufficient to
distinguish MS--MS binaries from single MS stars over a very wide
range of magnitude, so this conclusion is tentative.

Individual stellar magnitudes and positions, and the detailed results
of the artificial-star experiments, are available from the author.

\section{Stellar Distributions in M30}

\subsection{The Cluster Center}

The position of the center of M30 is not known to within an accuracy
of several arcseconds, as was made clear by the near-mistargeting of
our images.  We need an accurate position for the center in order to
measure the cluster's surface-density profile.

The cluster center is usually taken to be a point of maximum symmetry
of the stellar density; thus, an accurate measurement of its position
requires an image of a large surrounding region.  We cannot find a
point of symmetry from our FOC images alone, since the general area in
which the center is located appears near their edge.  However,
Yanny\markcite{yanny} \etal\ (1994) took a set of \hst/WFPC2 images of
the M30 cusp in April 1994.  Their Planetary Camera (PC) image covers
a larger $20'' \times 20''$ area around the cluster center, although
the images are not as deep as our FOC images.

We obtained the Yanny \etal\ $V$-band PC image from the \hst\ Archive,
and used it to measure an improved position for the center as follows.
We first reduced the image using DAOPHOT and ALLSTAR and a rough
point-spread function.  We then selected a magnitude cutoff, guessed a
prospective center, and then found the number-weighted centroid of the
positions of the stars within some radius $r$ of that center.  We used
the result of that measurement as the next guess for the center's
position, and iterated until the position converged, which generally
took only two or three iterations.  We tried a variety of magnitude
cutoffs ($V = 18$ to 19) and radii $r$ (5 to $10''$).

The position of the center measured with various combinations of
parameters varied by $0\secspt5$ or so.  We took a typical position in
the middle of the range as our adopted cluster center.  We then
transferred its position to a position on the FOC image using a
transformation derived from the positions of several bright stars, and
finally to standard coordinates using the STSDAS {\sl xy2rd} task.
The position of the center, in J2000 coordinates, is
$(21\supr{h}40\supr{m}22\spt164, -23^{\circ}10'47\secspt12),$ and is
marked on Fig.\ 1.\footnote{This position corresponds to $(497,310)$
on the $V-$band PC image and $(50.4,195.3)$ on our $V-$band central
FOC pointing (the upper half of Fig.\ 1).  These two images are
available from the \hst\ Archive; their rootnames are U2AS0301T and
X2L50201T, respectively.}  The new position of the center is
$2\secspt6$ from the position given by Djorgovski \&
Meylan\markcite{table2} (1993b), which was the position used in
targeting the central FOC field---thus explaining the near-miss of the
center.  The uncertainty in the position of the center is at least
0\secspt5.

\subsection{The Surface-Density Profile}

The surface-density profiles of stars in the inner $12''$ of M30 are
shown in Figure 3, and are also given in Table 1.  Points connected by
lines in Fig.\ 3 refer to corrected counts, and lines without points
refer to uncorrected (raw) counts.  The stars are divided into three
magnitude bins.  The brightest bin (solid lines) contains stars in the
range $18.75 < V < 20.25$ (corresponding to $0.76 \Msun > M > 0.69
\Msun$, according to BV92).  The middle bin (dotted lines) contains
stars with $20.25 < V < 21.75$ ($0.69 \Msun > M > 0.57 \Msun$), and
the faintest bin (dashed lines) contains stars with $21.75 < V <
23.25$ ($0.57 \Msun > M > 0.49\Msun$).

The density measured in each radial and magnitude bin was corrected
for incompleteness using the matrix method of Drukier\markcite{dru88}
\etal\ (1988, see Paper I)\footnote{As a check, we also tried
correcting for incompleteness with a simpler method that ignored
bin-jumping (i.e., we simply counted the fraction of artificial stars
in a bin that were recovered, regardless of their output magnitude).
The results were completely consistent with what is shown in Fig.\
3.}.  Points are plotted in Fig.\ 3 only when at least half of the
artificial stars in the given bin were recovered; the middle magnitude
range is thus usable only for $r > 2''$, and the faint range only
outside of $3''$.

As expected, the density of stars in the brightest magnitude range
rises going into the cluster center, with a power-law slope near
$-0.8.$ The density profile of the middle group is very similar.  The
profile of the faintest bin appears to have a somewhat flatter slope,
although the errors are large enough for this last bin that its slope
cannot be distinguished from the other two.

The innermost points of the brightest bin are consistent with either a
continuation of the density cusp into the very center of the cluster,
or with a small core of radius $< 2''.$ In Paper I, we found that the
small number of turnoff stars in the central $\sim1''$ of M15 made it
impossible to tell the difference between a cusp and a small core.
With less than half the number of stars here, the distinction is even
more difficult.

\subsection{Maximum-Likelihood Profile Fitting}

As in Paper I, we used a maximum-likelihood method to fit our star
counts to parameterized density profiles of the form
\begin{equation}
f(r;\alpha,r_c) = {1 \over {[1 + C_\alpha (r/r_c)^2]^{\alpha / 2}}},
\end{equation}
where $\alpha$ is the asymptotic logarithmic cusp slope (for $r \gg
r_c$), and $r_c$ is the core radius (the radius at which the projected
density reaches half its central value; the constant $C_\alpha \equiv
2^{2/\alpha} - 1 $ makes this so for any reasonable $\alpha$).

Since the star counts are not complete, and the incompleteness varies
with position in the image, the observed counts must be fit to a
distribution that has incompleteness incorporated into it.  (Fitting
profiles to the corrected counts is not a good procedure, for reasons
explained in Paper I.)  So, the counts were actually fit to the
product $g(r) = f(r) \psubm(r),$ where $f$ is one of the functions
defined above, and $\psubm$ is the probability that a star will be
detected by our analysis procedure.

The profile fit requires values of $\psubm$ at the radius of each real
star.  We found these values from the database of artificial stars by
the following procedure (which is explained in more detail in Paper
I):\ We begin by considering the annular region defined by the 1250
artificial stars nearest in radius to the star in question.  (Near the
cluster center, this region is actually circular.)  Within that
region, we use the Drukier matrix method to correct the local observed
LF for incompleteness.  The ratio of the appropriate magnitude bins in
the local observed and corrected LFs gives us the value of $\psubm$ at
the radius in question.  We then smooth $\psubm$ to remove noise that
leads to problems in the fitting.

The completeness fraction $\psubm$ is shown in Figure 4 for each of
the three magnitude ranges given in the previous section.  (The
remaining fluctuations in $\psubm$ are a result of insignificant local
variations in density within the annulus in which $\psubm$ is
computed; they do not affect the profile fit.)  The brighter two
magnitude bins are more than 90\% complete over most of the field; the
completeness of the brightest drops to $\sim65\%$ in the central
$2''$, while the second-brightest falls more quickly.  The
completeness of the faintest bin fluctuates between 60\% and 80\%.
Along with the scatter in its binned density profile (Fig.\ 3), the
larger fluctuations in this group's completeness indicate that there
are simply too few stars in the group to measure its profile within
the cusp with confidence.

In Figure 5, we show the likelihood function for the brightest group
as a function of $\alpha$ and $r_c$.  The point of maximum likelihood
lies at $\alpha = 0.70$ and at small $r_c$.  As in M15, models with
core sizes up to $\gtrsim 1''$ fit nearly as well as the pure-cusp
model; the 95\% upper limit on $r_c$ is $1\secspt9$.  The 1-sigma
uncertainty of $\alpha$ is 0.08, and its 95\% confidence limits are
$\pm 0.18$.

In Figure 6, we show the likelihood functions for the brightest and
the middle group, as a function of the cusp slope $\alpha$ only.  The
core radius $r_c$ was held fixed at $1''$ for this computation, and
stars within $1\secspt6$ of the center were not used, so that the same
radial range could be used for both groups.  Because of these small
differences from the parameters used in making Fig.\ 5, the best-fit
cusp slope for the brightest group is $0.76 \pm 0.07$ (1-sigma) in
Fig.\ 6, vs.\ $0.70$ in the previous case.  The middle group's
best-fit cusp slope is slightly steeper:\ $0.82 \pm 0.11$.  Note that
the cusp slopes of these two groups are consistent with each other.

\subsection{Comparison with the $r=4\minspt6$ Field}

Next, we look at the cluster cusp as a whole, with the goal of
comparing its overall mass function (MF) with a MF measured farther
out in the cluster.

The completeness-corrected luminosity function (LF) is shown in Figure
7, for two radial ranges within the central cusp ($0'' < r < 6''$ and
$6'' < r < 12''$).  (These are the same data as are plotted in Fig.\
3, but the star counts are now binned in magnitude within wider radial
ranges, rather than the reverse.)  The LFs are tabulated in Table 2.
Next, we convert the magnitudes to masses, using the BV92 isochrone
plotted in Fig.\ 2.  The resulting MFs are plotted in Figure 8, along
with the MF measured at $r = 4\minspt6$ (very close to the cluster
half-mass radius) by Piotto, Cool, \& \markcite{pck97}King (1997),
using the WFPC2.  The MFs are also given in Table 3.

Note that the cusp LFs and MFs shown in Figs.\ 7 and 8 cannot be taken
to represent the LF or MF at any single radius, since they are an
average over the oddly-shaped observed area.  Model predictions should
be integrated over the same area in order to be compared with these
data.  To allow future modelers to do this integration, the corners of
the observed area, relative to the cluster center, are given in Table
4.\footnote{The corners of the M15 fields from Paper I may be obtained
from the author.}

The two cusp MFs shown in Fig.\ 8 are very similar, except for the
lower overall density farther out in the cusp.  This observation
reinforces the conclusion that the cusp slopes for the two mass groups
presented in \S3.3 are consistent with each other.

A comparison with the $r=4\minspt6$ MF, on the other hand, shows
strong mass segregation, in the sense expected from two-body
relaxation, and also as expected from the observations of
\markcite{bol89}Bolte (1989).  One way of making a direct and
quantitative comparison of the three MFs is to fit a power law to
each, over the range in mass where they overlap, and compare the
power-law slopes.  (Since the MFs are not well fit by power laws, this
comparison is rather rough.)  Over the range from $0.50$ to 0.76
$\Msun$, the best-fit slope for the $0''<r<6''$ MF is $-3.1 \pm 0.5$;
for the $6''<r<12''$ MF it is $-3.8 \pm 0.7$, and for the
$r=4\minspt6$ MF it is $1.9 \pm 0.6$, where the Salpeter slope is
1.35.  (The $r=4\minspt6$ value is particularly sensitive to the
choice of the low-mass cutoff in the fit.  For this reason, its slope
here differs from that given for M15 in Paper I, despite the fact that
Piotto\markcite{pck97} \etal\ [1997] found the two MFs to be
indistinguishable.)  Over the indicated mass range, then, the
outermost MF differs from either of the inner two at the $6 \sigma$
level.  This comparison is crude, but confirms that the outer and
inner MFs differ dramatically.

\section{King--Michie Modeling}

As a first step towards modeling M30, we used our data to construct a
King--Michie model of the cluster, as we did in the previous paper for
M15.  King--Michie models have a lowered-Maxwellian distribution
function (Michie \& Bodenheimer \markcite{mb63}1963, King
\markcite{irk66}1966), which approximates the steady-state solution of
the Fokker--Planck equation (King \markcite{irk65}1965).  These models
do not incorporate all of the important physical effects, most
notably, the deviation from the lowered-Maxwellian DF late in core
collapse (Cohn \markcite{hc80}1980).  However, they are among the
simplest models that can give a prediction of the variation of the MF
from the center of the cluster to its edge---in fact, a fairly
realistic prediction, if their assumption of equipartition is correct
(see the next section).

We used the Gunn \& Griffin\markcite{gg79} (1979) formulation of the
multimass King--Michie model.  We calculated the model by first
choosing core and tidal (limiting) radii of 0.04 pc ($= 1\secspt1,$ at
a distance of 7.5 kpc) and 28.5 pc ($= 13\minspt2$), respectively, to
agree with observation.  (The value of $r_c$ is somewhat arbitrary,
but any value less than $2''$ leads to a nearly identical model.)
Next, we defined sixteen mass groups, whose numbers of stars and
average masses were constrained to agree with the MF observed by
Piotto\markcite{pck97} \etal\ (1997) at $r=4\minspt6$.  (The measured
densities were actually rounded off slightly, to make the plotted MFs
more readable.)  We then added a group of 0.55 $\Msun$ white dwarfs,
chosen (again, somewhat arbitrarily) to contain 20\% of the cluster
mass.  Finally, we added 1.33 $\Msun$ dark remnants (1.4\% of the
cluster by mass), whose number and mass were chosen to make the
central density cusp agree with the observed surface-density profile.
The model velocities are isotropic at all radii; since we found a good
fit to the observed surface-density profile (SDP) with an isotropic
model, we saw no reason to add velocity anisotropy.

A comparison of the model with observation is shown in Figure 9, and
the model parameters are summarized in Table 5.  The lower right and
upper left panels of Fig.\ 9 show model quantities that were
constrained to agree with observation; these panels confirm that the
iteration procedure converged.  The lower right panel shows the model
MF at $r=4\minspt6$, and the MF measured at that radius by
Piotto\markcite{pck97} \etal\ (1997).  The upper left panel shows the
surface-brightness profile (SBP) of M30 given by Trager, King, \&
Djorgovski\markcite{tkd95} (1995), and the density profile of stars
with $18.75 < V < 20.25$ from Fig.\ 3.  The subpanel just below the
upper left panel shows the logarithmic difference, on an expanded
scale, between the SDP of group 1 of the model and the observed SBP.

The lower left and upper right panels show tests of the model, since
the quantities plotted in those panels were not constrained to agree
with observation.  The lower left panel shows the velocity-dispersion
profile (VDP) of the giants (the $0.78\Msun$ group).  The solid jagged
line is the VDP measured by Gebhardt\markcite{geb95} \etal\ (1995),
who used a nonparametric method to turn their individual velocity
measurements into a profile.  The dashed lines above and below this
line are the 90\% confidence band on their VDP.  The heavier, smooth
line is the model-predicted VDP for the same stars.  King--Michie
models are isothermal in their center, so the model here does not
reproduce the rise in velocity dispersion in the inner arcminute of
M30.

The upper right panel shows the two FOC mass functions from Fig.\ 8,
along with the MFs predicted by the model (which have been integrated
over the actual observed field).  Note that the agreement of the
observed MFs with the model MFs at the high-mass end is a result of
the fit of the surface-brightness profile, but the low-mass part of
the MF was not constrained in any way.  The model agrees with the
observed MFs at the low-mass end quite well.

\section{Discussion}

For M30, our two main results are as follows: (1) Within the central
cusp, there is very little mass segregation.  (2) The overall MF of
the cusp, though, is quite different from the MF at $4\minspt6$, and
the difference is consistent with the predictions of a King--Michie
(KM) model.  In M15, on the other hand (Paper I), a comparison of the
overall cusp MF with the outer MF revealed that their difference was
smaller than a KM model would predict.  Now, our next step is to
determine what these mass-segregation measurements imply about the
dynamical states of the two clusters.

This task is made more difficult by the fact that the KM models do not
fit the Gebhardt \etal\ velocity-dispersion profiles.  That lack of
agreement implies that the KM models also do not have the correct
cluster potentials and distribution functions (DFs); thus, we might
not expect them to predict the correct degree of mass segregation,
either (in which case our model comparisons would be meaningless).  An
alternative approach would be to run a set of computational
simulations, such as Fokker--Planck or $N$-body models.  Those
evolving models, though, often have the ability to fit {\it too}
easily, in the sense that decent fits can be found at one model time
or another for many different sets of input parameters.  In that case,
it becomes difficult to determine which of the physical effects are
most important.

Instead, here we will take a more direct approach, by trying to
determine what our data imply about the current state of the cluster
{\it without} assuming that the DF has any particular form, or that
its present form must be the result of a specific set of physical
processes.  A number of recent globular-cluster studies have adopted a
similar philosophy; see, for example, Merritt \& Tremblay
\markcite{mt94}(1994), Cote \etal\ \markcite{cote95}(1995), or
Gebhardt \& Fischer \markcite{gf95}(1995).

What, then, can observations of mass segregation tell us about the
structure of a cluster?  Suppose that we somehow knew the form of the
cluster potential, say, by inferring it from a large number of
measurements of giant-star velocities.  In a known potential, the
observed degree of mass segregation reflects the allocation of energy
between various stellar species, since groups of stars with lower
({\it i.e.}, more strongly bound) energies will tend to be found
deeper in the potential well.  The classical view is that a cluster
should evolve towards equipartition of energy, in which two stellar
species $a$ and $b$ would have velocity dispersions $\disp{a}$ and
$\disp{b}$ such that $m\subr{a}\disp{a} = m\subr{b}\disp{b}$ at the
cluster center.\footnote{Actually, this condition is correct only for
systems in which the DF is strictly Maxwellian.  The lowering of the
Gaussian in the King--Michie DF implies that KM models are not truly
in equipartition (Pryor, Smith, \& McClure \markcite{psm86}1986).  A
more general condition for equipartition is that the rate of exchange
of kinetic energy between species is zero; for more details, see \S8.3
of Binney \& Tremaine \markcite{bt87}(1987).  However,
high-concentration KM models do satisfy the simpler condition given
above quite closely, near their centers.}  However, core-collapse
simulations have shown that equipartition is usually {\it not}
achieved.  In fact, as Spitzer (1969) first showed, the initial
attempt to approach equipartition leads to a runaway contraction of
the group of high-mass stars, and this ``equipartition instability''
leads to rapid core collapse in any cluster with a realistic mass
function (Inagaki \& Saslaw \markcite{is85}1985).

Now, if a cluster {\it is} in equipartition, lower-mass stars will
tend to have higher velocities at the center than higher-mass stars,
so they will tend to reach larger radii, and will have a flatter
density profile.  On the other hand, if the species are not in
equipartition, then the relative density distributions will differ
from the equipartition prediction:\ for example, if one species has
velocities that are ``too high,'' its density profile will be flatter
than it would be in equipartition.  The current spatial distributions
of stellar species thus reflect the ability of stellar encounters to
transfer energy between species over the course of the cluster
history, and give us a view not only of the present structure of a
cluster, but also of its past.

Our present stumbling block, though, is our lack of knowledge of the
cluster potential.  However, we shall see below that having a {\it
detailed} model of the potential is less important than it may
initially seem.

\subsection{The Cluster Envelopes}

For the moment, let us consider the comparison of the central MFs with
the outer MFs in each of the two clusters, and leave the mass
segregation within the central cusps for later.

\subsubsection{Mass Segregation as an Indicator of Equipartition}

As described above, if the KM-model potentials had resulted in
velocity-dispersion profiles that fit the observations, then the
agreement (for M30) or disagreement (for M15) of the model-predicted
mass segregation with observation would have directly indicated
whether the relevant stellar species were in equipartition.  Without a
well-fitting model, we must find another way to infer what our
observed densities imply about the stellar velocities.

Since the relaxation time at any radius in a globular cluster is much
longer than the crossing time at that radius, we can approximate the
cluster as being nearly collisionless.  For our purposes here, let us
also assume that each cluster has an isotropic velocity ellipsoid at
all radii, and that neither of the clusters rotate.  The density and
velocity profiles should then obey the isotropic non-rotating Jeans
equation, which in spherical coordinates is (see \S4.2.1[d] of Binney
\& Tremaine \markcite{bt87}1987):
\begin{equation}
M(r) = - {{r v_{j}^2} \over G} \Biggl(
{{d \ln \nu_{j}} \over {d \ln r}} + {{d \ln v_{j}^2} \over {d \ln r}}
\Biggr),
\end{equation}
where $M(r)$ is the total mass enclosed within radius $r$, and
$\nu_{j}$ and $v_{j}^2$ are the spatial density profile and the radial
component of the velocity-dispersion profile, respectively, of species
$j$.  For two species $j$ and $k$, the ratio of radial velocity
dispersions at any radius must then be
\begin{equation}
{v_{j}^2 \over v_{k}^2} = 
{({{d \ln \nu_{k}} / {d \ln r}}) + ({{d \ln v_{k}^2} / {d \ln r}})
\over
({{d \ln \nu_{j}} / {d \ln r}}) + ({{d \ln v_{j}^2} / {d \ln r}})}.
\end{equation}

In principle, given the density profiles of two species and the VDP of
one species $j$ (say, the giants), Eq.\ 3 would allow us to solve for
the unobserved VDP of the other species $k$ (say, a group of
lower-mass stars).  Our interest here, though, is not in the VDP of
the lower-mass stars itself (since we do not have enough information
to determine it with much confidence), but in the value of the
velocity ratio $\vv{j} / \vv{k}$ near the cluster center.  If we can
find a way to use the available data to infer this value, then we will
know whether the species are in equipartition, {\it i.e.}, whether
they satisfy the condition $m_j\vv{j} = m_k\vv{k}$.

Below, we will describe how we used the observed densities (of both
giants and lower-mass stars) and velocities (of giants) in each
cluster to infer the velocity ratio given by Eq.\ 3.  In the end, it
will turn out that this procedure will confirm the impression given by
the King--Michie models in the previous section:\ the data presented
in this paper are consistent with M30 being in equipartition, while
the data from the previous paper show that M15 is not.  Before
continuing the discussion of our method, it is worth asking why the KM
models give nearly the right answer on the equipartition question,
despite the fact that they fall short in other ways.  The answer to
this question will provide some insight into the relative importance
of density and velocity measurements in constraining different aspects
of cluster structure.

To see the reason for the KM models' success, note that the observed
VDP slopes ($d \ln v^2 / d \ln r$) are much smaller than the
density-profile slopes ($d \ln \nu / d \ln r$) over most of each
cluster's radial range.  Now, consider a hypothetical set of cluster
models that all have the same density profiles for species $j$ and
$k$.  The VDPs (and thus the potential and distribution functions) may
differ from model to model, but those differences change the value of
the numerator and denominator of Eq.\ 3 by very little; thus, all
models within the set must have nearly the same velocity ratio.  In
other words, while the VDP of one particular species may differ from
model to model, the {\it ratio} of the VDPs of a pair of species {\it
cannot} differ greatly between models.

To be more specific, imagine that we had a model of, say, M30, that
used a non-King--Michie DF.  Suppose further that this model was a
correct and complete representation of the cluster structure, in the
sense that it reproduced all of the available data:\ not only would it
fit the density profiles of both the giants and the lower-mass stars
(as did our KM model above), but also the VDP of the giants (whose
slope was not fit by our KM model).  Now, since this non-KM model and
our KM model have identical density distributions for the giants and
identical density distributions for the low-mass stars, they must be
members of a set of the type defined above, and Eq.\ 3 then requires
that the velocity ratios of the two models be similar.  But, since our
KM model is in equipartition, the non-KM model must then be close to
equipartition as well---and, therefore, so must the cluster itself!

What we have shown (in an approximate way) is the following:\ If the
observed {\it densities} in a cluster (such as M30) are well fit by
the density distributions of a KM model fit to that cluster, then the
cluster must be fairly close to being consistent with {\it all}
predictions of some other non-KM model that is also in equipartition.
Conversely, if a cluster is far from matching KM-model
mass-segregation predictions (e.g., M15), then {\it no} equipartition
model can adequately describe the cluster, since all such models would
make similar mass-segregation predictions (by an argument similar to
the one given above).  Thus, even though the KM DF was originally
derived without a knowledge of many of the physical processes that
actually take place in a cluster, KM-model {\it fitting} nevertheless
remains a valuable part of cluster-structure studies, since the models
make useful mass-segregation predictions, and are much easier to
calculate than more sophisticated models.

Our approach here, then, will be to use our data to infer the run of
the velocity ratio $\vv{j} / \vv{k}$ with $r$ that pertains to each
cluster, along with an estimate of its uncertainty.  If this inferred
run of the ratio is consistent with the run of the ratio from the KM
model, and approaches $m_{k} / m_{j}$ near the cluster center,
then we may conclude that the stellar species are in equipartition, to
within our measurement errors.  On the other hand, if the run of
$\vv{j} / \vv{k}$ found by the method is consistently and
significantly larger (or smaller) than the run of $\vv{j} / \vv{k}$ of
the cluster's KM model, then the lower-mass stars (species $k$) must
have velocities that are lower (or higher) than the equipartition
velocities.

\subsubsection{Detailed Procedure}

In practice, there are several difficulties that we must overcome if
we wish to use Eq.\ 3 to find $\vv{j} / \vv{k}$ as a function of $r$:\
(1) We have the {\it projected} density profile of the giants, not the
spatial profile.  (2) For the lower-mass stars, we have even less
information:\ only a few points on the projected profile.  (3) We also
see the VDP of the giants in projection, and it is noisy.

We can get around these difficulties, at least in part, using the
following methods, which take advantage of the fact that we have
already fit KM models to the clusters:

(1) {\bf The densities of the giants.}  
For the spatial-density profile of the giants (group $j$), we use the
giants' spatial-density profile from the KM model of the cluster in
question.  (Recall that the projection of this spatial profile was fit
to the observed surface-brightness profile.)  Although it is possible
that the deprojection of the observed profile is nonunique, the
possible range of the spatial profile should be small, since the
profiles are always rather steep.  (For a steep spatial profile,
line-of-sight effects are less important than the profile slope in
determining the shape of the projected profile.)

(2) {\bf The densities of the lower-mass stars.}
For the density profile of the group of lower-mass stars (group $k$),
we choose a density profile from the KM model that fits the projected
profile of that group at the observed radii.  For M30, since the
model-predicted mass segregation fits the observations, we simply use
the actual KM-model profile for the lower-mass group $k$.  For M15, we
take a different profile $l$ from the model:\ one whose radial ratio
of projected density ($\Sigma$) between two of our M15 fields outside
the cusp, $\Sigma_{l,{\rm model}}(20'') / \Sigma_{l,{\rm
model}}(4\minspt6)$, matches the observed $\Sigma_{k,{\rm obs}}(20'')
/ \Sigma_{k,{\rm obs}}(4\minspt6)$ of group $k$.

Admittedly, this matching of the density ratio is a very crude way of
``fitting'' a profile to two density measurements.  We have no way of
knowing how well the radial derivative of the selected profile $l$
matches the actual $d \ln \nu_k / d \ln r$ of the lower-mass group
$k$, but without further observations, we have little choice.

(Note that this use of the model density profile does {\it not} imply
that our results are dependent on the KM distribution function or
potential.  The model density profiles are used only as fitting
functions, since King--Michie profiles are known to fit globular
clusters well.)

(3) {\bf The velocity-dispersion profiles.}
Since the observed VDPs of the giants are noisy, we do not actually
use them; in fact, we could not even evaluate Eq.\ 3 without
deprojecting the observed VDP.  Rather than attempting a deprojection,
we take the following alternative approach:\ We solve Eq.\ 2 for the
VDP of both the giants ($\vv{j}$) and of the lower-mass stars
($\vv{k}$), using a method to be described shortly, which will require
us to assume a potential for the cluster.  We then verify that the
projected VDP of the giants that we have calculated is consistent with
observation; if not, we adjust the potential and repeat the process
until the projected VDP is adequate.  The ratio of the calculated VDPs
of the giants and of the lower-mass stars then gives a result
equivalent to evaluating Eq.\ 3 directly.

The method used to solve Eq.\ 2 for $v^{2}(r)$ for each of the two
mass groups, given an assumed $M(r)$ (equivalent to a potential), was
as follows:\ For each group, we began by setting $d \ln v^2 / d \ln r
= 0,$ and using Eq.\ 2 to solve for $v^2$ as a function of $r$.  We
then took a radial derivative of the computed $v^{2},$ and used that
derivative as the $d \ln v^2 / d \ln r$ term in Eq.\ 2 again.  We then
proceeded to alternate between computing $v^{2}$ and $d \ln v^2 / d
\ln r$ until both converged.  The procedure did not converge in the
central part of the cluster, where numerical inaccuracies in the
density-profile derivative were amplified by the iteration procedure.
Over most of its radial range, though, $v^2$ converged after a few
iterations for both mass groups.  We could then take the ratio of
$v^{2}$ for the two groups, to get the desired velocity ratio.

In the first step of this procedure, we had to assume a potential.
For reasons given earlier, the final velocity ratio is only weakly
dependent on the exact form of the assumed potential; the only
important constraint is that the projected VDP be consistent with the
observed velocities of giants.  Thus, we chose to assume as simple a
form for $M(r)$ as was necessary to generate VDPs that were consistent
with the Gebhardt \etal\ measurements:\ a power law.  The only
parameters in the $M(r)$ fit were the power-law exponent itself (which
always corresponded to potentials that were steeper than an isothermal
sphere), and the total mass within the cluster limiting radius.  We
found the appropriate values of the parameters by trial and error:\ we
simply tried different parameter combinations until the projected VDP
of the giants agreed with the Gebhardt \etal\
observations.\footnote{Note that all of our knowledge of the potential
comes from the Gebhardt \etal\ data; our mass-segregation measurements
give us no new information about either the potential or the form of
the distribution function.}

\subsubsection{Results}

The results of using this Jeans-equation procedure on the M15 and M30
data are shown in Figure 10.

The lower panel on each side of Fig.\ 10 shows the line-of-sight
velocity dispersions of giants obtained by iteratively solving Eq.\ 2,
as well as the Gebhardt \etal\ VDPs for each cluster.  For M15, we
used a potential $\propto r^{-0.2},$ taking the power law from the
result of the nonparametric analysis of M15 velocity data by Gebhardt
\etal\ \markcite{geb97}(1997).  For M30, we found a good fit to the
velocities with a potential proportional to $r^{-0.6}.$

The solid lines in the upper panels of Fig.\ 10 show the ratio
$\vvr{giants} / \vvr{low-mass}$ for both clusters, as calculated using
Eqs.\ 2 and 3.  For M30, the ``low-mass stars'' are a group with
masses near $0.61\Msun$.  For M15, the low-mass stars have masses near
$0.52\Msun$, but in the iteration procedure we used a density profile
for $0.62\Msun$ stars from the Paper I KM model, as discussed above.
(Also, for M15 we used an isotropic KM model, rather than the
anisotropic model given in Paper I, since our method here requires the
assumption of velocity isotropy.  The isotropic and anisotropic models
make virtually identical mass-segregation predictions.)

The several solid lines shown in the upper panels of Fig.\ 10 are the
velocity ratios found after various numbers of iterations (from 2 to
5).  The curves separate from each other in the central cusp, where
the solutions have not converged, but are quite consistent for
$r>10''$.

The dashed lines in each upper panel show the velocity ratio (for the
same pair of species) predicted by the KM model of each cluster.
Since the KM model is (nearly) in equipartition, the dashed line
approaches the mass ratio $m\subr{low-mass}/m\subr{giants}$ near each
cluster center.  At larger radii, the KM-model velocity ratio
approaches 1, as the tidal cutoff becomes more and more important.

For both clusters, the velocity-ratio curves found from the Jeans
equation lie above the KM-model curve, implying that the low-mass
stars have slightly lower velocities than they would if the two
species were in equipartition ({\it i.e.}, the two species have
velocity dispersions that are closer together than they would be in
equipartition).

(Note that any other mass group could have been taken to represent the
lower-mass stars in each cluster, with similar results.  The exact
values of $\vvr{giants} / \vvr{low-mass}$ would have been different,
of course, but the solid curves lie above the dashed curves no matter
which low-mass group is used.)

In interpreting Fig.\ 10, it is important to realize that we have
inferred the run of the velocity ratio from a {\it small} number of
measurements of the densities of low-mass stars in each cluster, not
from a measurement of their entire density profile.  Since we used
KM-model density profiles as fitting functions in the iteration
procedure, it should come as no surprise that our inferred
velocity-ratio runs resemble those of the KM model as well, except
that they are shifted in the vertical direction.  The proper way to
interpret each of the upper panels, then, is not to pay attention to
the details of the radial variation of $\vvr{giants} /
\vvr{low-mass}$, but instead to regard each group of solid curves as a
{\it single} measurement of their offset from the KM-model curve.

The crucial question, then, is whether those offsets are significantly
different from zero.  The two major sources of uncertainty in our
Jeans-equation procedure are the errors in the MFs themselves, and our
lack of detailed knowledge of the radial dependence of the potential.
For both clusters, an estimate of the errors shows that the MF error
dominates, and leads to an uncertainty of $\sim0.06$ in $\vvr{giants}
/ \vvr{low-mass}$ at the innermost value of $r$ for which the several
curves plotted in each upper panel remain together before separating
in the central cusps.

Within this formal error of $0.06$, the M30 Jeans-equation curve is
consistent with the KM-model curve for that cluster, while for M15,
the Jeans-equation curve is $\sim2.5\sigma$ away from the KM model.
We conclude, then, that over our observed ranges in mass and radius in
each cluster, the stellar species in M30 are in a state of
equipartition of energy, while those in M15 most likely are not.

There are some potential sources of systematic error that we have
neglected in this analysis.  The most important of these is the
possibility of velocity anisotropy in either of our clusters.  An
anisotropic velocity ellipsoid would lead to the presence of a third
term on the RHS of Eq.\ 2.  In that case, it would be more appropriate
to determine the radial run of the ratio of $(v^2\subr{r} +
2v^2\subr{t})$, rather than the ratio of just $v^2\subr{r}$.  We will
not pursue such a computation here, but will merely note that strong
anisotropy could seriously undermine our equipartition conclusions,
particularly if the degree of anisotropy varied with radius, or
especially if it differed for different stellar species.

This last possibility may not be so far-fetched.  Takahashi (1996,
1997) and Takahashi \etal\ (1997) have investigated the development of
anisotropy in Fokker--Planck models of multimass clusters, and have
found that the degree of anisotropy can differ widely from model to
model, and from species to species within a single model.  Since there
is no observational data that could constrain the shape of the
velocity ellipsoid in either cluster, we can only consider the
isotropic case, and leave the possibility of anisotropy as an open
question.  (The last of the three Takahashi papers finds that the
velocity ellipsoid becomes more isotropic after core collapse, so for
these two clusters, our assumption of isotropy may be fairly close to
the truth.)

Our assumption of non-rotation might also be questioned, at least in
the case of M15, where Gebhardt \etal\ (1997) have detected a small
degree of rotation.

\subsection{The Cluster Cusps}

In principle, the same sort of Jeans-equation analysis could be
applied to the data in the central $\sim10''$ of all three clusters.
However, the small number of stellar velocities measured in the cusps
does not allow us to constrain the potentials well; in fact, it is not
known with certainty whether any clusters harbor central massive
objects.  However, we {\it can} make some reasonable assumptions about
the potential, and ask what our data would then imply.

First, though, we compare the data with some simple models from the
literature that take equipartition, rather than a particular
potential, as a basic assumption.

\subsubsection{Simple Equipartition Models}

Post-core-collapse and black-hole models of cluster cusps in
equipartition predict very different variations of cusp slope with
mass, since the two models have greatly differing distribution
functions.  In a simple equipartition core-collapse model (Cohn
\markcite{hc80}1980), the potential is proportional to $r^{-0.23}$,
and there is a strong degree of mass segregation within the cusp,
which can be described as follows:\ if $-\beta$ is the logarithmic
slope of the spatial density cusp for a group of stars with mass $m$
(with the projected cusp having slope $\alpha = \beta - 1$, for $\beta
> 1$), then $\beta = 1.89 (m / m\subr{max}) + 0.35,$ where
$m\subr{max}$ is the mass of the stars that dominate the cusp (Cohn
1985).  Black-hole cusps, on the other hand, have much weaker
segregation (Bahcall \& Wolf \markcite{bw76}1976,
\markcite{bw77}1977); the potential is Keplerian ($r^{-1}$), and all
of the stellar species have spatial cusp slopes between $-1.75$ and
$-1.50$.

More specifically, then, in M30 the core-collapse model would require
the presence of dark stellar remnants with masses of $\sim1\Msun$ to
account for the turnoff stars' projected slope of $-0.76$.  This model
would then predict a projected slope of $-0.54$ for the ``middle''
($0.57\Msun$ to $0.69\Msun$) group in that cluster---a value
$2.5\sigma$ away from the measured slope of that group ($-0.82$).  It
has also been suggested that an isothermal sphere in equipartition
would be a good model of a cusp well after core collapse.  Such a
model, though, would also predict a projected slope near $-0.5$ for
the lower-mass stars, and would not match the data.

The similarity of the two cusp slopes in M30 {\it is} quite consistent
with the predictions of the black-hole model.  However, kinematic
evidence for a black hole in M30 is lacking.  A black hole would
induce a cusp in the velocity dispersion of giants, but Zaggia \etal\
\markcite{zag92}(1992, \markcite{zag93}1993) concluded that the
cluster is isothermal near its center, with $\disp{giants} = 6.0 {\rm
\ km\ s}^{-1}$.  The VDP estimated nonparametrically by
\markcite{geb95}Gebhardt \etal\ (1995; see Fig.\ 9) does rise in the
center ($\vvr{giants} \propto r^{-0.4}$), but more slowly than
Keplerian.  A Keplerian $v^2 \propto r^{-1}$ cusp would barely fit
within their 90\% confidence band, while an isothermal region like
that found by Zaggia \etal\ would fit easily.  Velocities have been
measured for only a few cusp stars, though, and it is {\it possible}
that a velocity cusp has been missed.  A $\sim1000\Msun$ black hole
might be detected in a globular cluster by the STIS spectrograph
recently installed on \hst.

\subsubsection{Non-Equipartition Models}

An alternative to comparing our cusp slopes with models from the
literature would be to use the procedure from \S5.1 once again, this
time using the observed cusp slopes to get values for $d \ln \nu / d
\ln r$.  One step in the procedure was to find a potential that fit
the observed velocities of giants; but with so few velocities for cusp
stars, we have no hope of constraining the cusp potential in this way.

Instead, we can simply assume the same crude power-law potential that
we fit to each cluster in \S5.1.  The density profile follows a power
law within the cusp; if we assume that the velocity-dispersion profile
$v\subr{r}(r)$ and potential $\Phi(r)$ also follow power laws, then
the Poisson equation requires that $d \ln v\subr{r}^2 / d \ln r = d
\ln \Phi / d \ln r$.  Thus, the process of finding the velocity ratio
$\vvr{giants} / \vvr{low-mass}$ within the cusp becomes a simple
matter of substituting the appropriate values into Eq.\ 3, rather than
iteratively solving Eq.\ 2 as we did previously.

For M15, this Jeans-equation procedure is especially attractive, since
the potential that best fit the Gebhardt \etal\ data was proportional
to $r^{-0.2},$ and thus closely resembles the cusp potential predicted
by Fokker--Planck core-collapse models.  If we use this potential and
our observed cusp slopes for the ``turnoff'' and ``middle'' groups
from Paper I ($-0.64$ and $-0.56$, respectively) in Eq.\ 3, we find a
value of $0.96 \pm 0.09$ for the ratio of squared central velocity
dispersions for these groups of $\sim0.8$ and $0.7 \Msun$ stars.
(Note that these are not the same groups that were used on the right
side of Fig.\ 10.)

In equipartition, the value of the velocity ratio for these two M15
groups would be $\sim0.87$.  Our star counts therefore do not indicate
a {\it significant} departure from equipartition in the cusp of M15,
although we do not go deep enough that this measurement is very
sensitive.\footnote{In Paper I, we showed that the M15 cusp star
counts were not well matched by one particular equipartition
core-collapse model, while here, we cannot say whether the cusp is in
equipartition.  The reason for the difference in conclusions is that
the uncertainty of the velocity ratio given here includes a
contribution from the uncertainty in the potential, which in turn
arises from the uncertainty in the velocity-dispersion profile.  The
comparison in the previous paper, on the other hand, did not involve
any velocity data---it was a comparison of density slopes only.}  It
is worth noting that since the publication of Paper I, Dull \etal\
\markcite{dull}(1997) have published a new Fokker--Planck model of
M15, which matches our observed cusp slopes fairly well.

In M30, we can use the cusp slopes given in \S3.3 for the ``turnoff''
and ``middle'' groups to find a velocity ratio of $1.03 \pm 0.08$
(assuming a potential proportional to $r^{-0.6}$, as we did in \S5.1).
The equipartition prediction for this ratio is $0.86$, so our
measurement is off by $2.1\sigma$ from this prediction.

Although our measurements do not entirely exclude the possibility of
equipartition in either cusp, the similarity of the two measured
slopes in each cluster is nevertheless quite intriguing.  The density
profile of each species within the cusp is a result of the behavior of
the distribution function of that species at the lowest ({\it i.e.},
most strongly bound) energies.  The observation that the cusp slopes
are similar therefore suggests that those portions of the distribution
functions are also quite similar, no matter what the potential might
be within the cusp, and regardless of whether a black hole is present.
However, the distribution functions of the various species cannot be
{\it entirely} the same, since the MFs do vary with radius outside of
the cusp.  Thus, we are left with the possibility that the physical
processes that govern the dynamics of the most strongly bound stars of
the cluster---and that therefore determine the structure of the
density cusp---may not be the same as the processes that determine the
structure of the rest of the cluster.  In the next section, we give
some more speculative ideas for what might be happening in the cusps.

\subsection{An Evolutionary Scenario}

Our main conclusions are as follows:\ (1) There are notable
differences between clusters in the allocation of energy between
stellar species, with some clusters, such as M30, being closer to
equipartition of energy than others, such as M15.  (2) In some cases,
the cusp may be a different environment from the rest of the cluster,
in the sense that there is less mass segregation within the cusp than
one would expect from measurements outside of the cusp.

Let us consider the second conclusion first.  A high binary fraction
in the cusp, or the presence of a central black hole, would lead to a
difference between the cusp and the envelope, or could account for
differences between cusps of different clusters.  Direct evidence for
these circumstances is lacking, but either is plausible.

Even in a cluster composed only of single objects of stellar size,
though, a concentrated population of heavy dark objects could account
for these results, if we allow for dynamical evolution.  Suppose that
a cluster begins with similar density and velocity distributions for
all stellar species, as a result of violent relaxation.  The
highest-mass stellar group will then undergo core collapse, as a
result of the Spitzer equipartition instability.  After core collapse,
the central density of this highest-mass group is much higher than
that of the present-day main-sequence stars, and the relaxation time
of the subsystem of high-mass objects in the core and cusp is
considerably less than the relaxation time in the surrounding
envelope.  The core and cusp might thus undergo rapid gravothermal
oscillations, confined to the central few arcseconds, and the
(relatively) few luminous stars in the cusp would simply react to the
fluctuating potential produced as a result of these oscillations.
What we observe in the cusp might then differ considerably, depending
on the phase of oscillation in which we happen to see it.

After the first core collapse, what happens to the stars that spend
most of their time in the cluster envelope?  Energy production from
binaries in the core has stabilized the cluster against further
collapse, and compared to internal relaxation, external effects on the
global MF operate rather slowly (unless the cluster suffers frequent
tidal shocks).  It is plausible that the stellar envelope can then
continue its dynamical evolution towards equipartition, on timescales
determined more by the overall properties of the cluster than by the
properties of the core.

We might expect, then, that the stars in a cluster would reach
equipartition at a given radius only when the cluster age becomes
equal to the relaxation time at that radius.  We can use our KM models
of M15 and M30 to evaluate the relaxation time for each cluster at
$r=4\minspt6,$ the radius of our outermost observations:\ for M30,
$t\subr{r} \simeq 10$ Gyr, while for M15, $t\subr{r} \simeq 40$
Gyr.\footnote{In this computation we have taken the mean density
inside of $r=4\minspt6$ as the density of the stars being encountered,
the overall mean stellar mass as the mean mass of the stars being
encountered, and the velocity dispersion at $r=4\minspt6$ as the
velocity of the star suffering encounters.  These quantities could be
chosen somewhat differently, since the relaxation time is only an
order-of-magnitude quantity, but the relaxation time for M15 would be
considerably longer no matter how we do the computation.}  So, the
$r=4\minspt6$ field in M15 is at a radius where $t\subr{r}$ is greater
than the cluster age, while relaxation at $r=4\minspt6$ in M30 should
be nearly complete---a scenario that is entirely consistent with our
observation that M30 is closer to equipartition than its more massive
cousin.

It is plausible, then, that that the data presented here and in the
previous paper represent two stages of the dynamical evolution of a
typical old and dense globular cluster.  Future observations of
cluster MFs, in a variety of clusters and in a number of fields within
each cluster, will tell us whether this scenario is close to the
truth.

\acknowledgements

The author would like to thank Ivan King for obtaining the images used
in this paper, as well as for many suggestions over the course of the
project.  He would also like to thank Jay Anderson and the referee,
Piet Hut, for careful and critical readings of the paper.  This work
was supported by NASA grant NAG5-1607.

\newpage

%
%

\def\capone{
F480LP (FOC ``$V$'') image of the two $7'' \times 7''$ FOC pointings
in M30.  The cluster center is indicated by a cross near the left edge
of the upper pointing.
}

\def\captwo{
A color--magnitude diagram, in Johnson $B$ and $V$, of the two FOC
fields in M30.  The dashed line is the main-sequence ridge line of
Bolte (1987), and the solid line is the ([Fe/H] $= -2.03,$ $t = 16$
Gyr) isochrone of Bergbusch \& VandenBerg (1992).  The redward
deviation of the FOC points from the line is due to the onset of
nonlinearity, which affects the $B$ magnitudes more strongly than the
$V$ magnitudes.
}

\def\capthree{
The projected stellar density in the cusp of M30, plotted in radial
bins, in three magnitude ranges.  The circular points connected by
lines refer to the counts corrected for incompleteness.  The lines
without points show the ``raw'' (incomplete) counts.
}

\def\capfour{
The completeness ratio $p_{\rm m}(r)$ for the three stellar samples in
M30.  See the text for an explanation of the fluctuations.
}

\def\capfive{
The log-likelihood of the ``bright'' M30 sample ($18.75 < V < 20.25$),
as a function of the (negative) cusp slope $\alpha$ and core radius
$r_c$ in arcseconds, using the fitting function defined by Eq.\ 1.
Lighter areas indicate a higher probability of seeing the observed
counts.  Contour lines are plotted one unit apart, so that each line
represents a factor of 10 in the likelihood function.  Thus, a model
with an $(\alpha,r_c)$ pair that lies on the $-2$ contour line is
one-tenth as likely to produce the observed counts as a model whose
$(\alpha,r_c)$ lies on the $-1$ line.  The point of maximum likelihood
lies at $\alpha = 0.7$ and very small $r_c$.
}

\def\capsix{
(Upper panel) The likelihood of seeing the observed M30 star counts as
a function of the (negative) cusp slope $\alpha$, for the brighter two
stellar samples, using the fitting function defined by Eq.\ 1.  The
solid curve refers to the brightest magnitude bin ($18.75 < V <
20.25$), and the dotted curve to the middle bin ($20.25 < V < 21.75$)
(Lower panel) The same likelihood functions converted to a cumulative
probability, assuming a uniform prior.
}

\def\capseven{
Luminosity functions in two radial ranges in M30.
}

\def\capeight{
Mass functions in two radial ranges in M30 (from our FOC images), and
a mass function farther out in the cluster (from the WFPC2 images of
Piotto, Cool, \& King 1997).  The two FOC MFs refer to the left-hand
scale, and the outer WFPC2 MF refers to the right-hand scale.  Since
the stellar density is much lower in the $r=4\minspt6$ field, the 
vertical position of the WFPC2 MF has been chosen to match one of the
FOC MFs at the bright end, so that the two may be easily compared.
}

\def\capnine{
The King--Michie model, fit to our observations and to others from
the literature.  In all panels (except the lower left), points are
observations, and solid lines are model constraints or predictions.
(UL) The surface-density profile (model constrained to
fit the observations).  The large inner points are from this work;
small outer points are from Trager \etal\ (1995).
(LL) The velocity-dispersion profile.  The narrow jagged line is the
profile produced by Gebhardt \etal\ (1995) from a nonparametric
analysis of their Fabry--Perot data.  The dashed lines are their 90\%
confidence band on their profile.  The heavier line is the profile
predicted by our King model, which does not reproduce the central rise
in velocity dispersion.
(LR) The mass function at $r=4\minspt6$ (model constrained to fit the
data), from Piotto \etal\ (1996).
(UR) Mass functions in the FOC fields, from this work (model matches
the observed MFs).
}

\def\capten{
The result of iteratively solving the Jeans equation (Eq.\ 2 in the
text).  Left panels refer to M30, and right panels refer to M15.
{\it Lower panel on each side:}\ Line-of-sight velocity dispersion of
giants, as found by solving the Jeans equation (solid lines), and as
measured by Gebhardt \etal\ (1994, on the left, with the profile
indicated by a dashed line, and its 90\% confidence limits indicated
by dotted lines), and by Gebhardt \etal\ (1995, on the right, indicated
by points).  The agreement with the observed VDPs shows that the
potential being used in the solution of the Jeans equation is
approximately correct.
{\it Upper panel on each side:}\ Ratios of velocity dispersions of
stars of differing mass, equivalent to either side of Eq.\ 3 (see
text).  The solid lines refer to the solution of the Jeans equation,
using different numbers of iterations.  The dashed lines show this
velocity ratio for the King models described in this paper and in
Paper I.
}

\ifthenelse{\boolean{figs}}{

%
%

%
%
\newpage

\begin{figure}
\centerline{\epsfxsize=10cm\epsffile{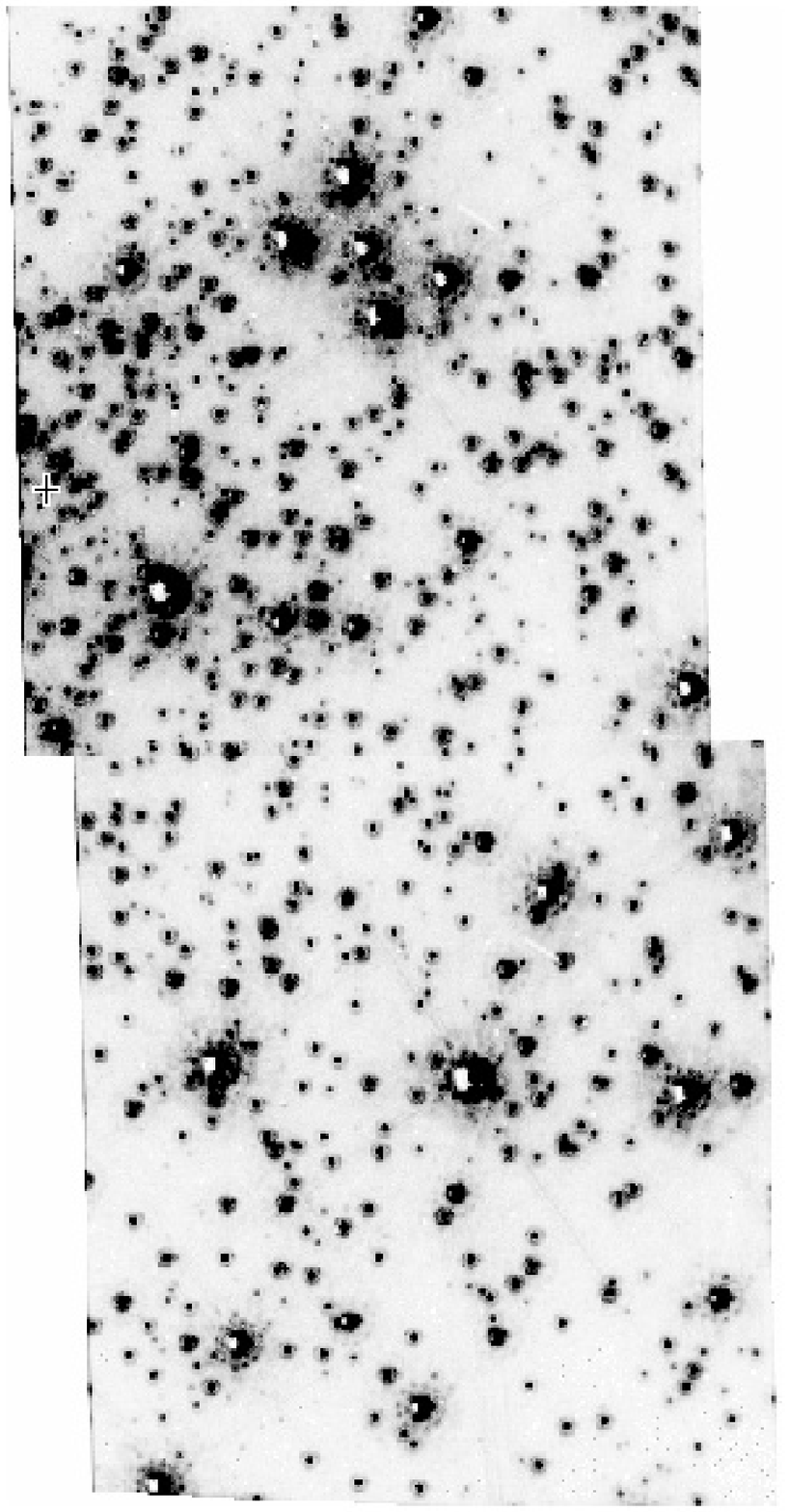}}

\vskip0.2truein
Figure 1. \capone
\end{figure}

%
%
\clearpage

\begin{figure}
\plotone{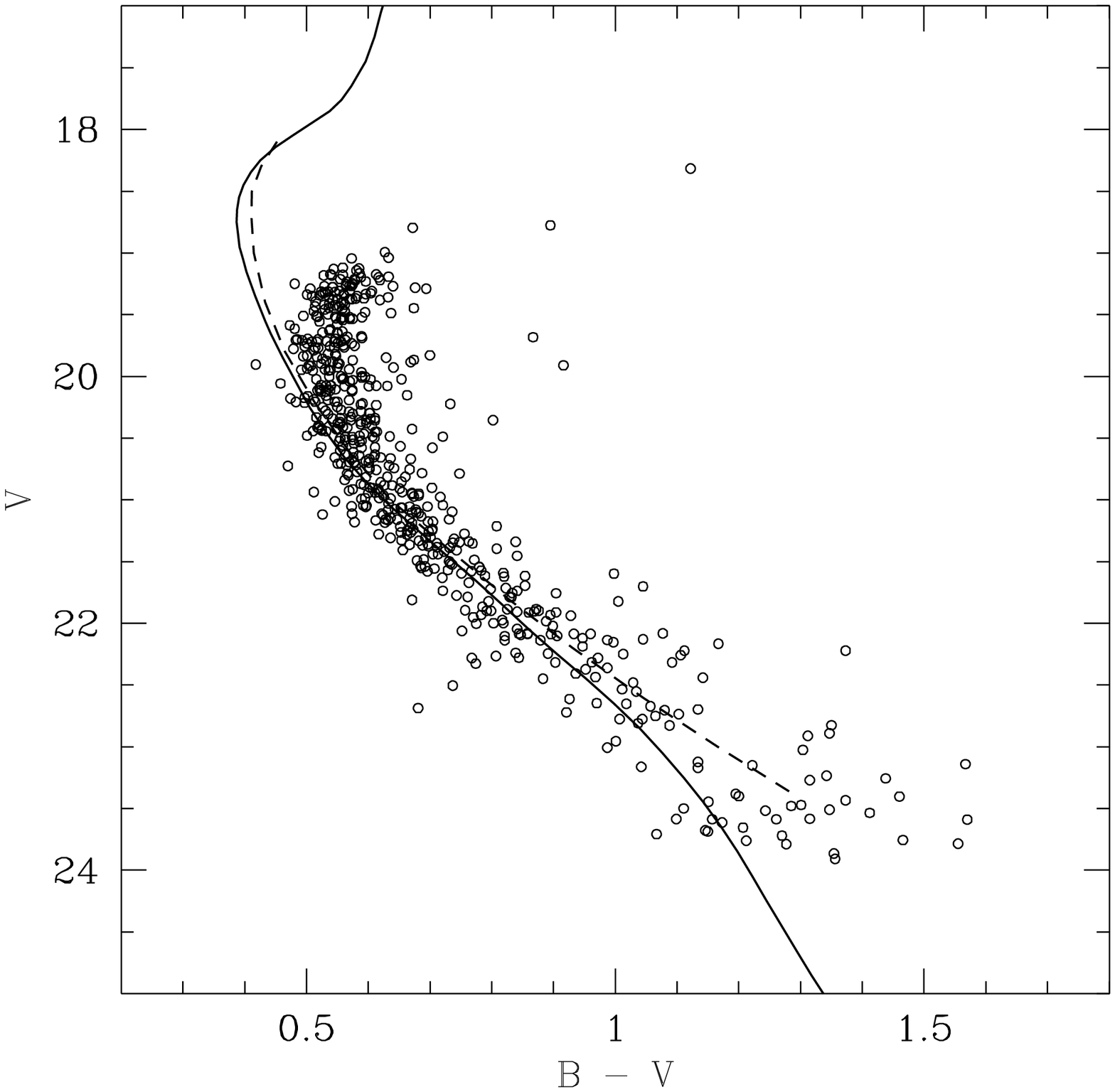}

Figure 2.  \captwo
\end{figure}

%
%
\clearpage

\begin{figure}
\plotone{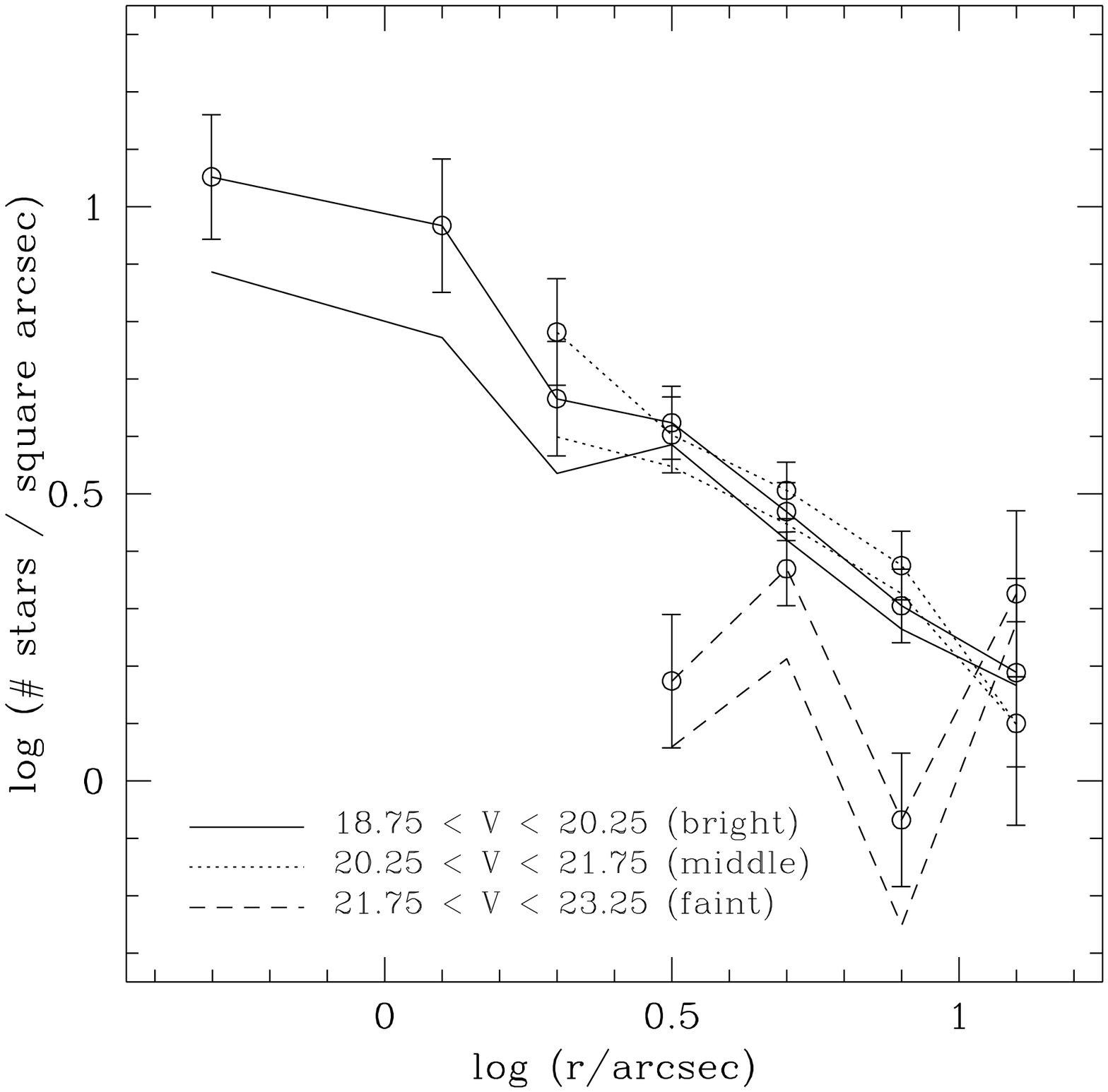}

Figure 3.  \capthree
\end{figure}

%
%
\clearpage

\begin{figure}
\plotone{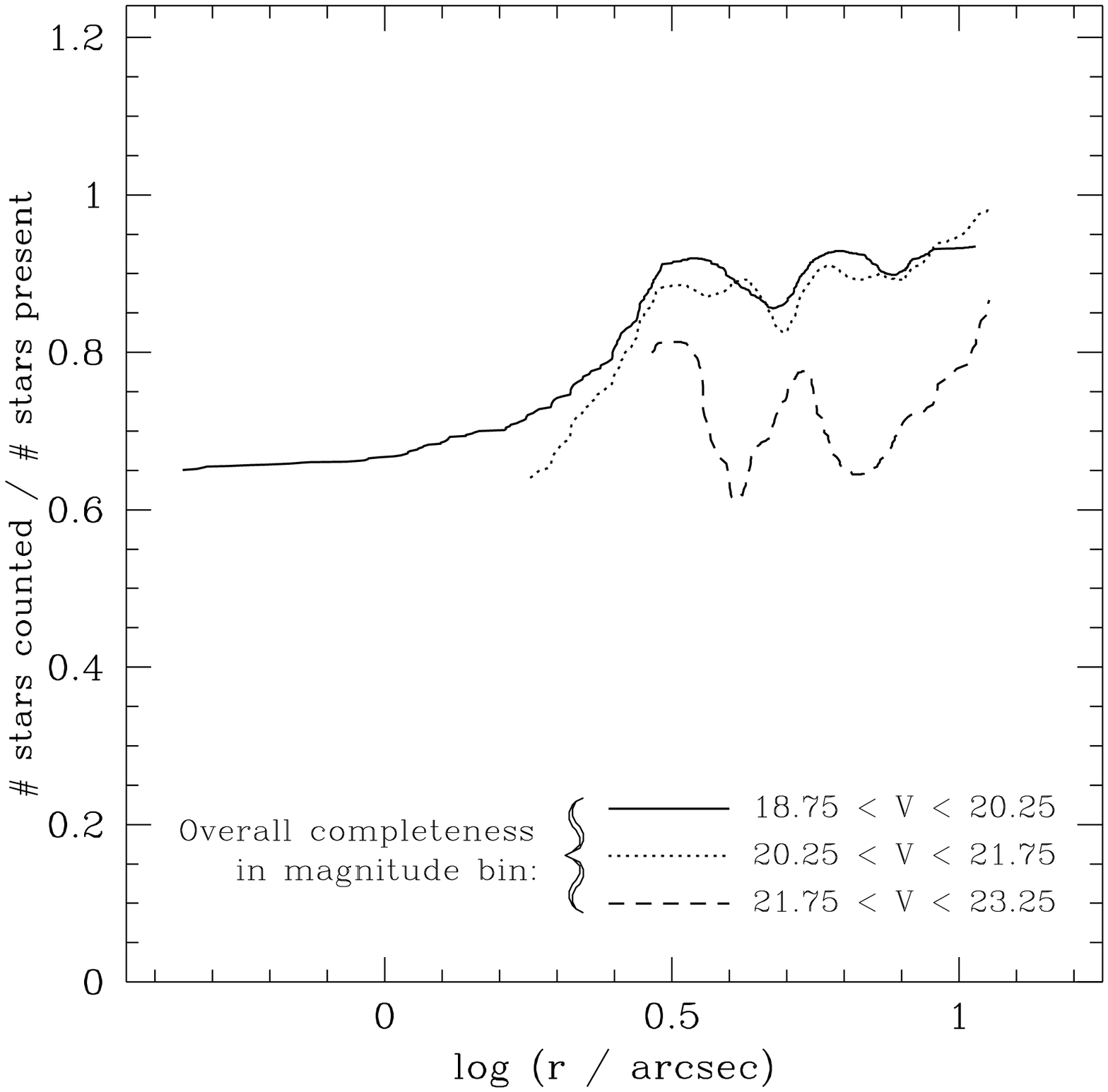}

Figure 4.  \capfour
\end{figure}

%
%
\clearpage

\begin{figure}
\plotone{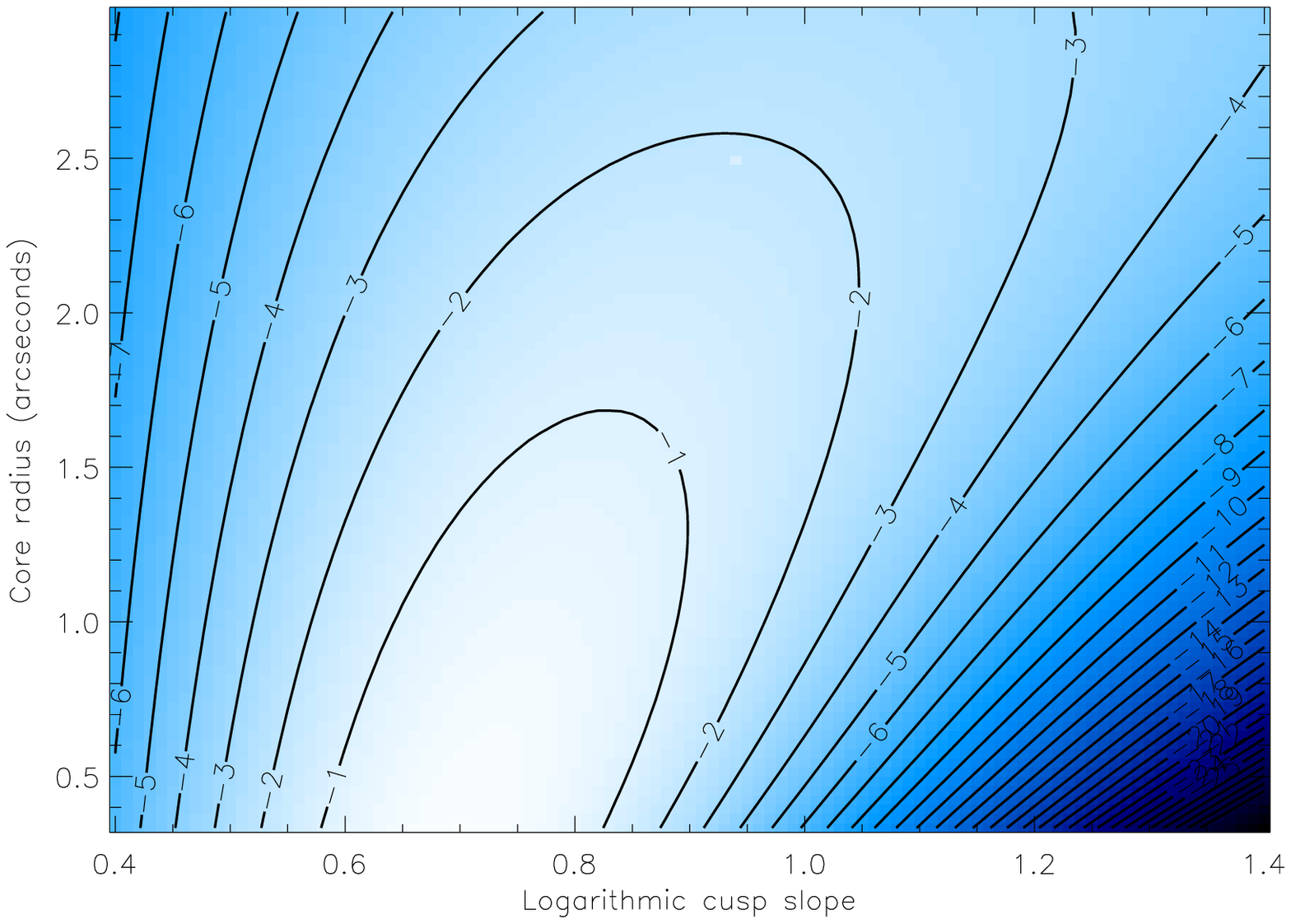}

Figure 5.  \capfive
\end{figure}

%
%
\clearpage

\begin{figure}
\plotone{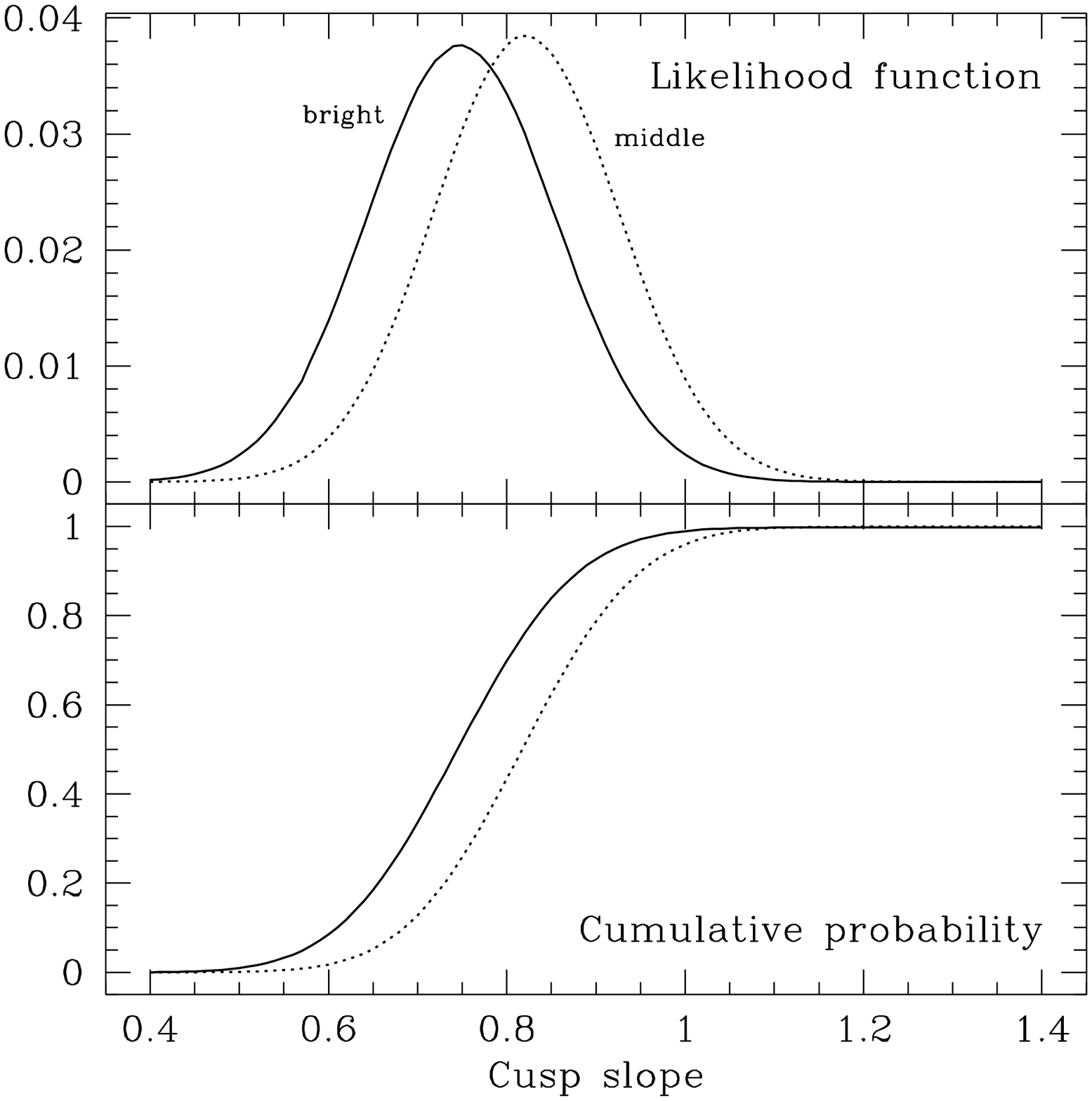}

Figure 6.  \capsix
\end{figure}

%
%
\clearpage

\begin{figure}
\plotone{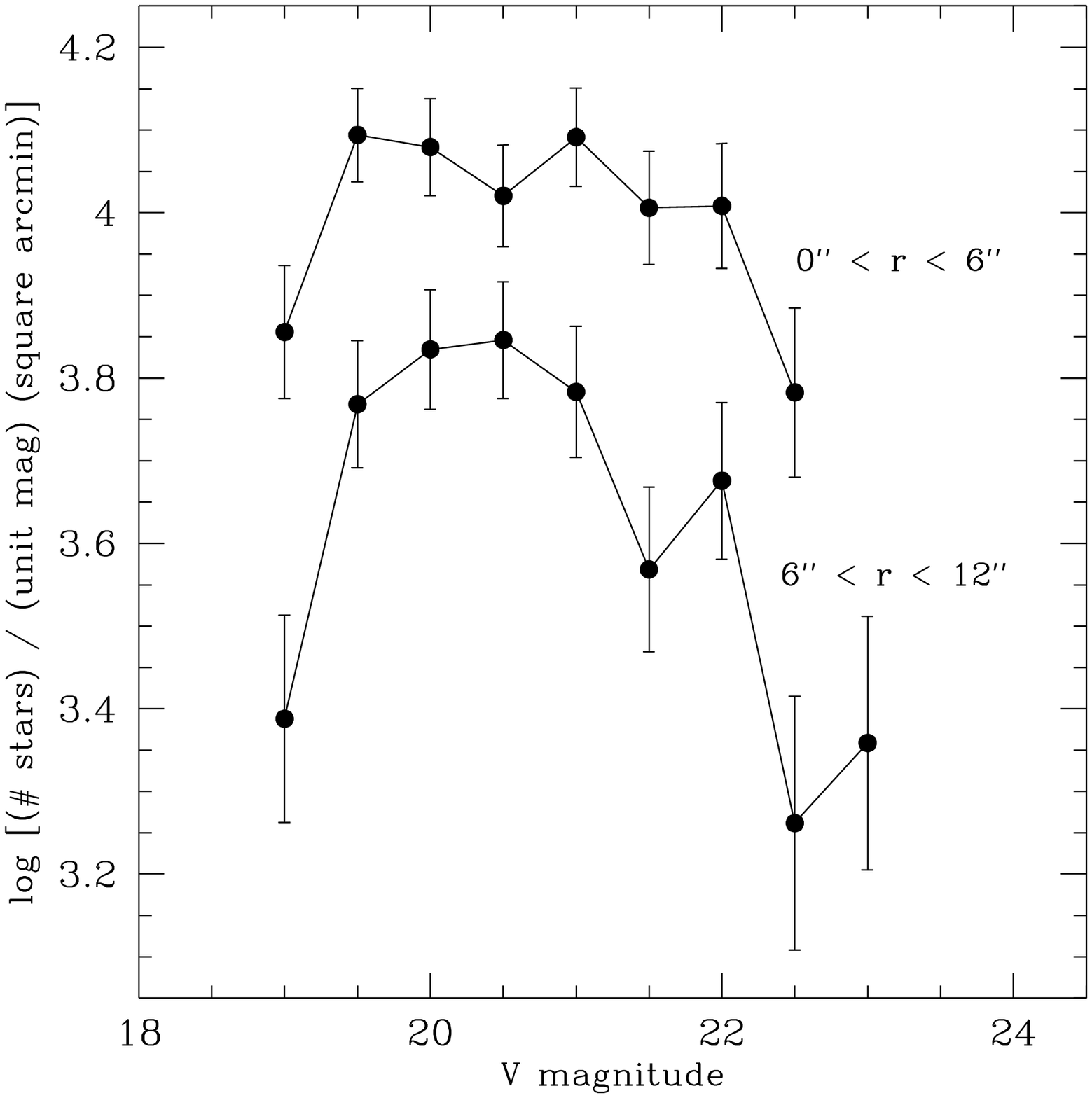}

Figure 7.  \capseven
\end{figure}

%
%
\clearpage

\begin{figure}
\plotone{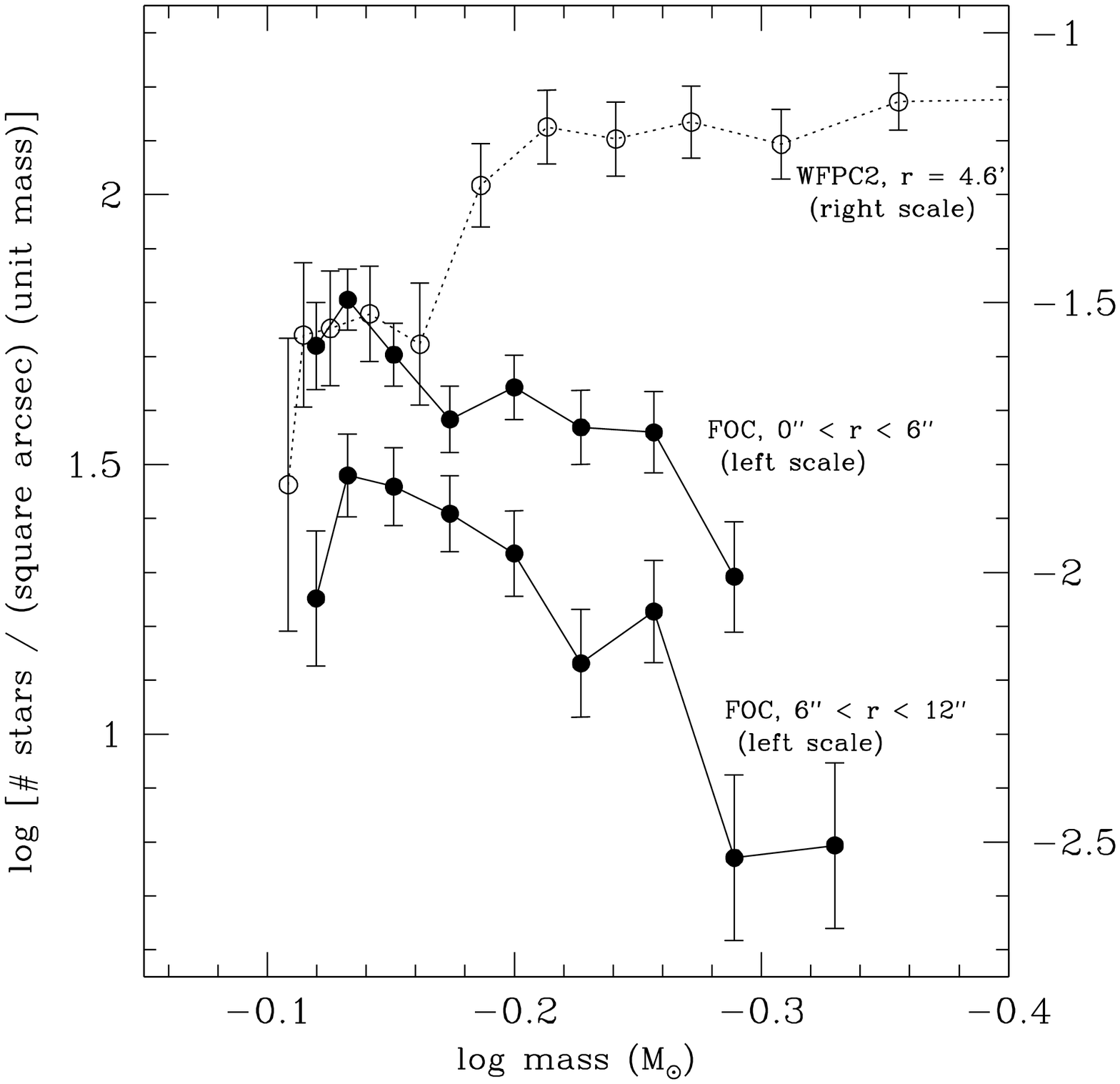}

Figure 8.  \capeight
\end{figure}

%
%
\clearpage

\begin{figure}
\plotone{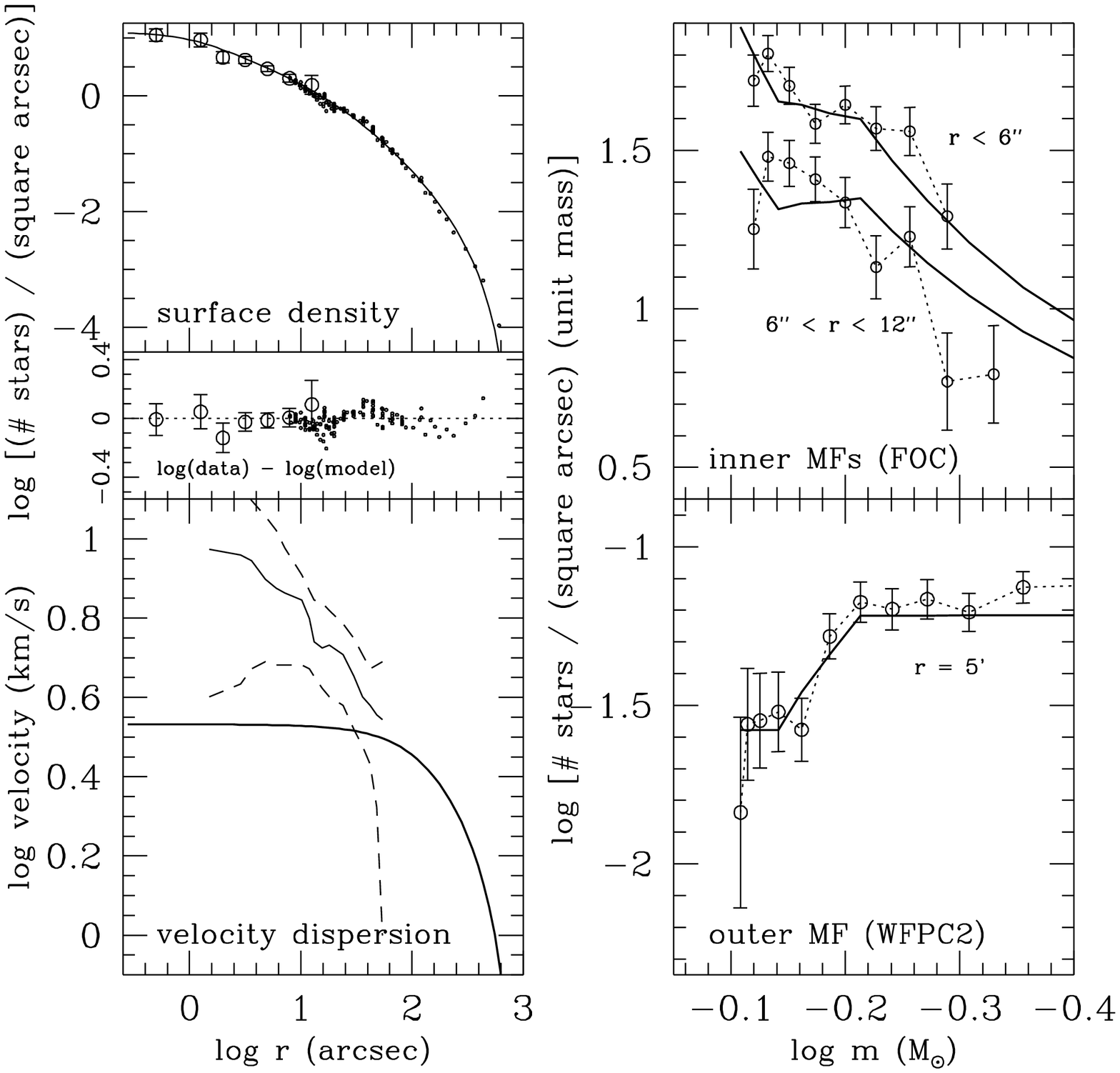}

Figure 9.  \capnine
\end{figure}

%
%
\clearpage

\begin{figure}
\plotone{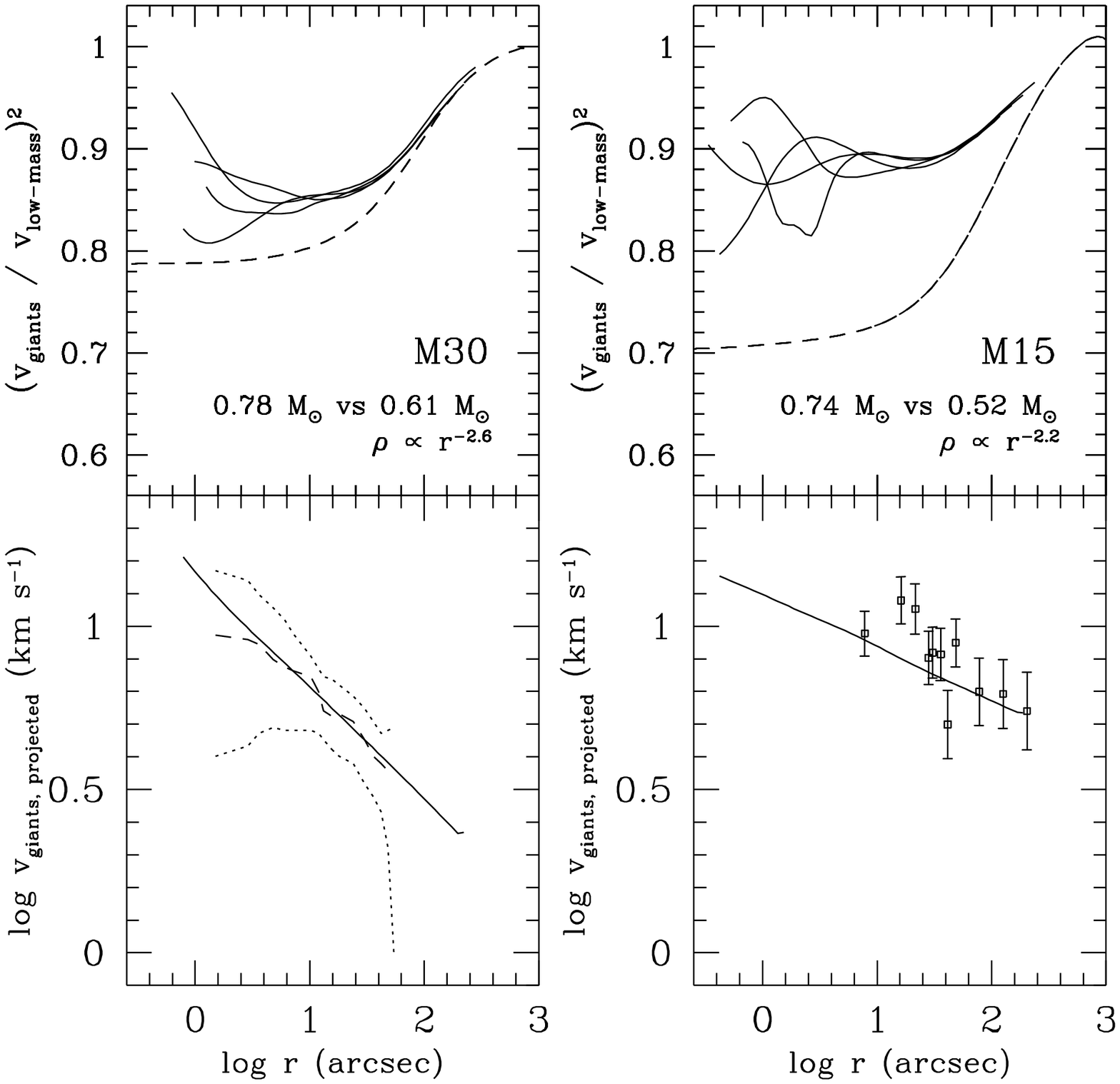}

Figure 10.  \capten
\end{figure}

}{

%
%

\figcaption[Sosin.fig1.ps]{\capone}

\figcaption[Sosin.fig2.ps]{\captwo}

\figcaption[Sosin.fig3.ps]{\capthree}

\figcaption[Sosin.fig4.ps]{\capfour}

\figcaption[Sosin.fig5.ps]{\capfive}

\figcaption[Sosin.fig6.ps]{\capsix}

\figcaption[Sosin.fig7.ps]{\capseven}

\figcaption[Sosin.fig8.ps]{\capeight}

\figcaption[Sosin.fig9.ps]{\capnine}

\figcaption[Sosin.fig10.ps]{\capten}

}

\end{document}